\documentclass[aps,preprint,prb,superscriptaddress,amsmath,showpacs,tightenlines]{revtex4}
\usepackage[dvipdfmx]{graphicx}
\usepackage{amssymb}

\begin{document}
\title{Physically Unclonable Function using Initial Waveform of Ring Oscillators}

\author{Tetsufumi Tanamoto, Shinich Yasuda, Satoshi Takaya and Shinobu Fujita}
\affiliation{Corporate R \& D center, Toshiba Corporation,
Saiwai-ku, Kawasaki 212-8582, Japan}

\date{\today}
\begin{abstract}
A silicon physically unclonable function (PUF) is considered to be one of the key 
security system solutions for local devices in an era in which the internet is pervasive.
Among many proposals, a PUF using ring oscillators (RO-PUF) has the advantage 
of easy application to FPGA. In the conventional RO-PUF,  frequency difference 
between two ROs is used as one bit of ID. 
Thus, in order to obtain an ID of long bit length, the corresponding number of RO pairs 
are required and consequently power consumption is large,  leading to difficulty in
implementing RO-PUF in local devices.
Here, we provide a RO-PUF using the initial waveform of the ROs. 
Because a waveform constitutes a part of the ID, the number of ROs is greatly
reduced and the time needed to generate the ID is finished in a couple of system clocks.
We also propose a solution to a change of PUF performance 
attributable to temperature or voltage change.
\end{abstract}

\maketitle

\section{Introduction}
In order to connect a huge number 
of devices as an Internet of Things (IoT), individual 
devices have to be manufactured at low cost.
In addition, to protect personal information, 
a stable security system must be implemented.
A physically unclonable function(PUF) is considered to be 
one of the important mechanisms for 
providing a unique ID for each device at low cost.
A PUF outputs a response ID  to a challenge signal.
The origins of the PUF signal  come from  process variations of transistors
and circuits.
Many types of PUF have been already proposed.
An arbiter PUF that uses the circuit delay is the most 
intensively investigated~\cite{arbiter,Lee}.
An SRAM-PUF that uses the initial memory value 
of SRAM is considered the most advanced~\cite{Guajardo,Holcomb}. 
Altera has announced its intention to implement an SRAM-PUF unit 
in the company's high-end FPGA.
Initial defects of memories~\cite{Marukame,RRAM} or 
trap sites in the transistor~\cite{Chen} can be also used to constitute a PUF. 
The first PUF based on ring oscillators (RO-PUF)  
was presented by Suh {\it et al}.~\cite{Suh},
in which the frequency difference between two ROs
is used as the PUF output.
The RO-PUF has an advantage in that it can be implemented 
in commercial FPGAs without difficulty and the stability is 
proved~\cite{Maiti0}.
However, previous RO-PUFs using the frequency difference 
between RO pairs have disadvantages: 
the number of RO pairs corresponds to 
the length of the ID, resulting that a large number of ROs is required. 
Thus, from viewpoint of IoT which requires low-power other than low-cost,
continuous running of RO of conventional RO-PUF is not desirable.

Because  FPGAs are widely used as the basic tools
to implement designs, FPGA-based PUFs are expected to 
protect the individual properties.
Here we propose a PUF based on a ring oscillator (RO) using 
the initial waveform of the RO output just after the oscillation starts.
We use an RO pair in which one RO is used for sampling of the other RO.
Because the frequency of RO ($\sim$ 1~GHz) is much faster than the system clock
($\sim$ 50-100 MHz), the sampling process is finished within a couple of 
cycles of the system clock, resulting in fast ID generation and 
low power consumption.
We tested our idea on Altera Cyclone V (28nm), and Xilinx (Spartan-6 and Spartan-3E) devices.
Our PUF is flexible because there is no design constraint. 
We use a simple Verilog circuit.

The rest of the paper is organized as follows.
In section II, we discuss related works regarding the RO-PUF.
In section III, we present our RO-PUF using a starting wave of the RO output.
Experimental details and results are presented in section IV.
Section V is devoted to the performance evaluation of the proposed  PUF. 
In section VI, we discuss temperature dependence.
Section VII presents conclusions. 

\section{Related works}
The mechanism of previous proposed RO-PUFs involves comparing 
frequencies between two ROs, and the large frequency variance 
leads to robustness of RO-PUFs.
In the first proposal of an RO-PUF~\cite{Suh}
the best RO pair with large frequency difference 
is selected out of many ROs, resulting 
in the requirement of a large number of ROs. 
Maiti {\it et al.} found that the RO frequency depends on 
the location on the FPGA, and improved the reliability of an RO-PUF 
by implementing each RO in one configurable logic block~\cite{Maiti1}.
Merli {\it et al}. show a chain-like structure that  
enhances the frequency difference using the frequency difference
 of nearest neighbor ROs~\cite{Merli}.
Habib {\it et al}.  proposed a more efficient RO-PUF by considering the 
programmable delay lines~\cite{Habib}.

There are many proposals regarding PUFs based on FPGA other than RO-PUFs.
Tuyls {\it et al}. proposed the butterfly-PUF~\cite{butterfly} and Flip-Flop(FF)-PUF~\cite{FF-PUF}.
The butterfly-PUF shows excellent performance of the Hamming distance 
and small temperature dependence.
The butterfly-PUF uses a bistable state of two FFs and 
is an FPGA version of the SRAM-PUF.
Although the butterfly-PUF shows excellent PUF performance, 
precise symmetry is required in the bistable FF structure. 
Yamamoto {\it et al}. proposed a PUF using RS-FF~\cite{Yamamoto}
in which inhibition input  of RS-FF is used to reflect the chip identity.
These PUFs require routing constraints or hard macro, resulting 
in lack of generality in the conventional use of FPGA, in which 
various applications and IPs are implemented at low cost.

The RO-PUF presented here requires neither routing constraint nor hard macro.
Circuits automatically implemented by FPGA software are measured in 
three kinds of FPGA devices. Various PUF performances are estimated.

\section{Proposed PUF using waveform of ROs}
The origin of a PUF is the process variation of transistors and circuits, 
and the previously proposed RO-PUFs show that
the process variations appear in the frequency of ROs.
The frequency difference among different ROs 
means that the period of one cycle between rising edges 
differs depending on each RO. 
Thus, it is natural to consider that the waveform difference of an RO
can be used as the source of a PUF.
The RO begins to oscillate when the supply voltage is applied
or the enable switch is on.
When the initial output value of the RO is 0, 
the time to first rising edge of oscillation signals differs
depending on each RO.

However, because the frequency of ROs ($\sim$ 1~GHz) 
is much higher than the clock frequency of the conventional circuit board ($\sim$ 50-100 MHz),
the conventional RO set up shown in Fig.1~(a) is insufficient to 
capture the oscillation signal.
In order to capture the oscillation signal of the RO at higher time resolution, 
we use other ROs as the sampling clock of FF.
Figure 1(b) is the minimum unit of the proposed PUF using the waveform of ROs (wRO-PUF).
Because the output frequency of Fig.1(b) is still high,
we need FFs to convert the higher-frequency signals to the lower-frequency digital data
 used in the other circuit.
Figure 2 is the basic unit of the wRO-PUF, 
where the number of FFs corresponds to the length of 
generated ID for this element.

Let us analyze the waveform of Fig.1(b).
Figure 3 shows the two possible patterns of the initial waveforms of the 
two ROs.
When $EN=1$, the ROs start to oscillate, and 
the initial value of ROs is set to 0 by the NAND element in Fig.1(b).
The time to the first rising edge differs depending on ROs.
When the period of the waveform of RO1 is longer than that of RO2 ($t_1>t_2$),
the initial value of the output in Fig.1(b) is 0.
When $t_1<t_2$, the initial value of the output in Fig.1(b) is 1. 
The variation of the oscillation period depends on the properties of the transistors 
included in the ROs. Thus this mechanism resembles that of the conventional RO-PUFs.
Moreover, because the waveforms are analog data, 
the relative difference $|t_2-t_1|/t_2$ between two waveforms is the origin of
new degrees of freedom when they are digitized.
Figure~\ref{doublewave} shows an example where the output signal (OUT) pattern of wRO-PUF greatly changes 
depending on the relative change $t_1/t_2$ even when we limit $t_1>t_2$.
If we take the first 32bits from these output signals, 
the number of the different bits between the two signals is 13,
which corresponds to the Hamming distance of the two data.

\begin{figure}[t]
\begin{center}
\includegraphics[width=8.0cm]{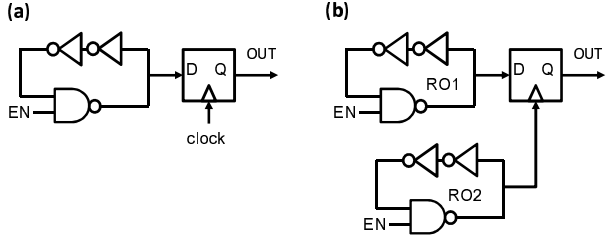}
\caption{ 
(a) Conventional PUF based on ring oscillators(ROs). The frequency of RO ($\sim$ 1 GHz) is much higher than the clock frequency of the conventional circuits ($\sim$ 50-100 MHz).
(b) Our proposed PUF consists of two ROs. In order to detect the high frequency output of RO1, 
the output signal of RO1 is sampled by the output of RO2.} 
\label{Start}
\end{center}
\end{figure}

\begin{figure}[t]
\begin{center}
\includegraphics[width=8.5cm]{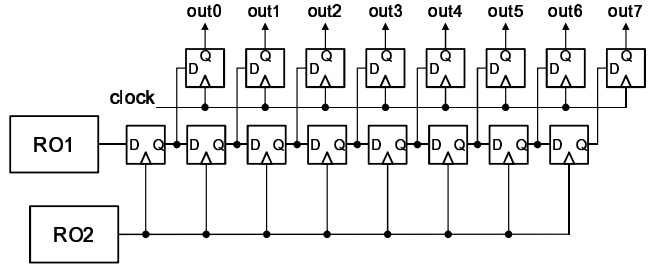}
\caption{ 
Our proposed PUF using the starting waveform 
of the RO. The waveform of the RO1 is captured by RO2. 
A shift register and output FFs are used to register the high frequency outputs of RO1.
The number of the outputs corresponds to an
ID unit length. In this figure, the length of output is 8 bit for 
simplicity.} 
\label{allcircuit}
\end{center}
\end{figure}
\begin{figure}
\centering
\includegraphics[width=6cm]{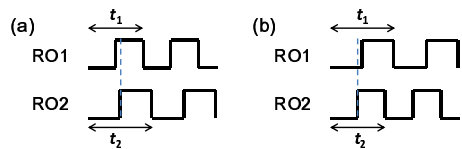}
\caption{
Two patterns of the initial signal of the two ring oscillators (ROs)} 
\label{waveform}
\end{figure}
\begin{figure}
\centering
\includegraphics[width=6cm]{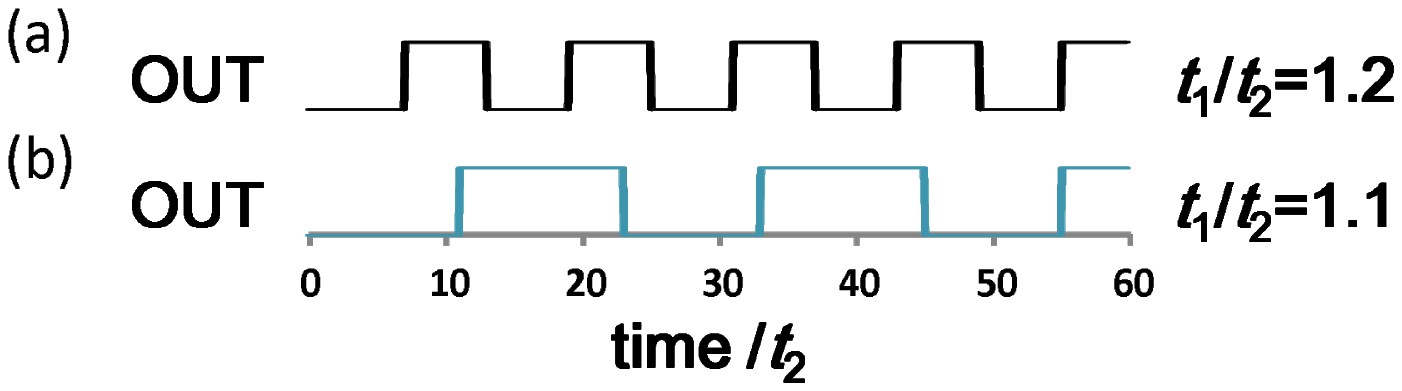}
\caption{
Simple simulations of output signals of wRO-PUF 
in Fig.1(b) when 
a frequency of RO1 is different of that from RO2.
(a) $t_1/t_2=1.2$ and (b) $t_1/t_2=1.1$.} 
\label{doublewave}
\end{figure}

We consider that the device ID of FPGA 
consists of outputs of several RO pairs in Fig.~\ref{allcircuit}.
Thus, the number of the FFs in Fig.~\ref{allcircuit} corresponds 
to the length of a unit ID component.
Because the frequency of RO is very high, it is considered that 
the sampling by ROs is out of the conventional operational region of FFs.
In particular, the change of signal in $'Q'$ terminal  
is compatible with $'D'$ terminal, 
resulting in the difference of output depending on the chip.
In addition, the output of Fig.~\ref{allcircuit} is 
considered to be affected by the wire delay between the output of RO2 and FFs.
This effect changes the periodic waveform to partly 
aperiodic waveform.
This mechanism resembles the arbiter-PUF or FF-PUF.
Thus, our wRO-PUF generates ID from the 
causes of RO-PUF (transistor variation) and  the arbiter-PUF(circuit variation).

\begin{figure}[t]
\begin{center}
\includegraphics[width=6.0cm]{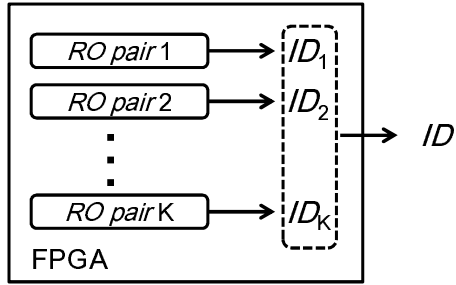}
\caption{ 
ID of each FPGA consists of several  RO pairs using waveforms. 
When each RO pair outputs
32 bit length ID, eight pairs $(K=8)$ generate 256-bit length ID of the FPGA. In this case, the number 
of ROs in an FPGA is 16 and the number of FFs is $32\times 2 \times 8=512$.} 
\label{ID}
\end{center}
\end{figure}

\begin{figure}
\centering
\includegraphics[width=8cm]{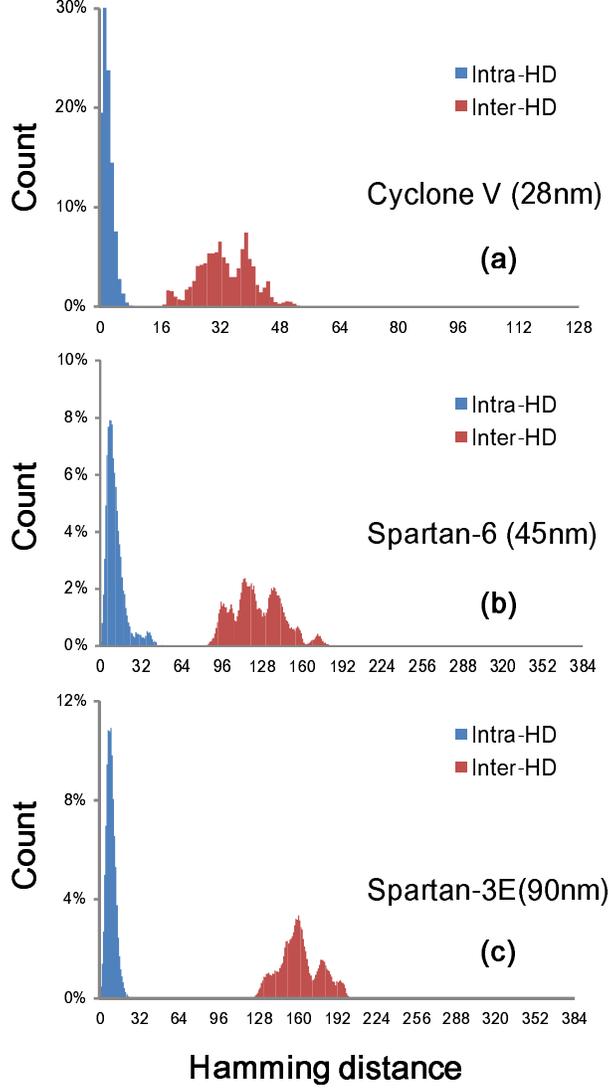}
\caption{
Hamming distances of the three FPGAs. 
In (a), the device ID  is composed of eight RO pairs of 16-bit length.
In (b) and (c), the device ID is composed of twelve RO pairs of 32-bit length.} 
\label{HD}
\end{figure}

\begin{table}
\begin{center}
TABLE I\\ 
{ Details of the datasets used.} 
\end{center}
\begin{tabular}{|c|c|c|c|}\hline
&  Cyclone V 
&  Spartan-6 
&  Spartan-3E 
\\ \hline
$N$:Number of devices
& 11 
& 11 
& 10 
\\
$K$: Number of RO pairs
& 8
& 12
& 12
\\
$ID_k$ length from one RO pair
& 16
& 32
& 32
\\
$T$: Number of tests per ID
& 1000
& 1000
& 1000
\\
System clock
& 100 MHz 
& 50 MHz 
& 50 MHz 
\\ \hline
\end{tabular}
\end{table}
\section{Experiments}
We apply our PUF to Xilinx Spartan-3E (xc6slx16-3csg324, Nexys2) and Spartan-6 (xc3s500e-4fg320, Nexys3)
and Altera Cyclone V GX (5CGXFC5C6F27C7N, Terasic).
ID is determined by the pattern of the wRO-PUF output that appears most frequently.
We described the circuit of Fig.2 by Verilog and used Nios II and Microblaze to 
control the communication between the FPGA devices and our personal  computer (Table I).
No design constraint is included.
Sampling is carried out by repeating $EN=0$ and $EN=1$, 
where the ID of the RO pair is the initial continuous $l_{\rm RO}$-bits after $EN=1$.

Examples of the measured waveforms of the five ROs of Spartan-6 
are given by
\begin{eqnarray}
{\rm RO \ pair}0: &011111111111000000000011111100000000, \nonumber\\
{\rm RO \ pair}1: &111111111000000001111111111100000000, \nonumber\\
{\rm RO \ pair}2: &000000011111111111100000000000000000,\nonumber\\
{\rm RO \ pair}3: &111110110101010110101010110100000000,\nonumber\\
{\rm RO \ pair}4: &000000000000000000001110000100000000,\nonumber
\end{eqnarray}
The waveforms of the five ROs show different bit patterns 
after the automatic placement and routing
reflecting the difference in  the RO frequencies 
depending on their positions~\cite{Maiti0}.
It can be seen that the waveforms include some irregular bit patterns that are not
inferred from the  Fig.~\ref{ID}. These irregular bit patterns are considered to 
be the results of wiring delay variances of the many-FF chains in Fig.\ref{allcircuit}.

The waveform after a long-time oscillation is 
affected by environmental noise.
Thus, long ID generation is undesirable 
and it is better to produce a long ID by combining outputs of several RO pairs. 
We define the ID of each device by the combination of 
output ID of $K$ RO pairs of the device as illustrated in Fig.~\ref{ID}. 
That is, when the 
output sequence of $k$-th RO pair defines $ID_k$, the total ID of the device 
is given by 
\begin{equation}
ID=\{ ID_1, ID_2, \dots, ID_K\}.
\end{equation}
The total ID length, $L$, is given by $L=l_{\rm RO} \cdot K$.

The most important metric to estimate a PUF performance is 
the Hamming distance (HD), which is the number of 
different bits between two outputs.
The waveform data are obtained by repeating the process of $EN=1$ and $EN=0$.
Here, the ID is defined as the pattern that appears most frequently over $T=1000$ experimental data. 
The intra-chip HD is the HD between the ID and all sampled data in the same chip after repeated measurements.
The count of the intra-HD ideally has a peak at zero HD.
The inter-chip HD is defined by the HD between different chips, 
and is ideally distributed around half of the ID length.
Figure \ref{HD} shows  HDs for the experimental results of the 
three kinds of FPGAs. 
It is seen that the intra-HDs are distributed around their most frequent ID patterns.
The peaks of the inter-chip HDs in Fig.~\ref{HD} are smaller than the half of the ID length.
The gap between the peak of the intra-HDs and those of the inter-chip HDs 
becomes smaller as the transistor size decreases.
This gap is smallest for the Cyclone V FPGA.
As mentioned above, the PUF properties of our wRO-PUF originate from 
the variations of RO frequency and wiring delay. 
As the technology node shrinks,  
the parasitic capacitance of the wiring is reduced, 
resulting in the decrease of delay variances.
It is considered that the reducing variations of the wire delay 
reduces the gap between the intra-HDs and the inter-HDs. 

\begin{figure}
\centering
\includegraphics[width=7.5cm]{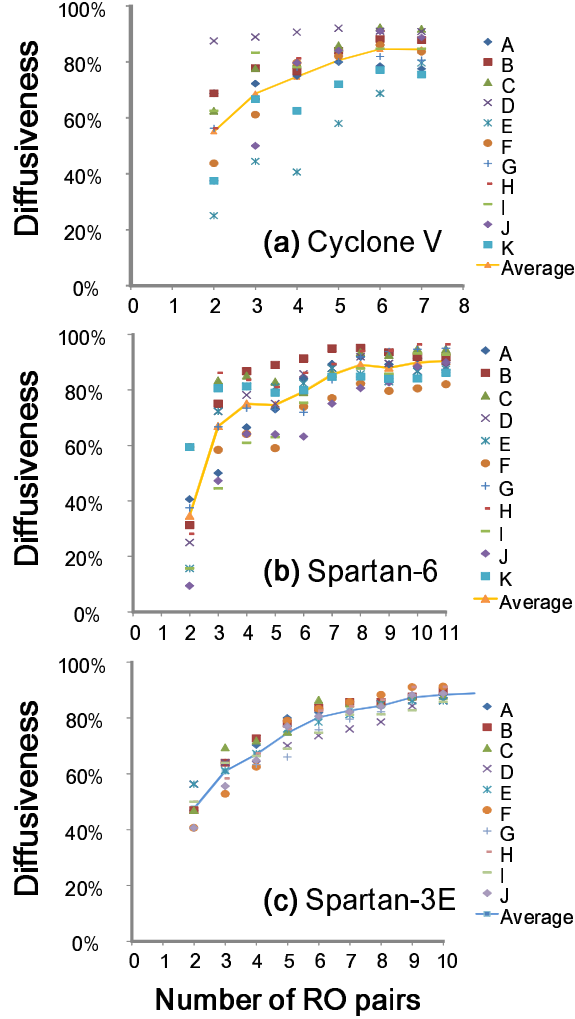}
\caption{
Diffusiveness defined in Ref.\cite{Hori} for 
the three types of devices.
In (a) eight RO pairs are measured. In (b) and (c), twelve RO pairs are measured.
$A$, $B$, $C$... show the FPGA device names} 
\label{diffusive}
\end{figure}

The ID distribution over many RO pairs can be 
measured by the diffusiveness introduced by Hori {\it et al.}~\cite{Hori}:
\begin{equation}
{\rm Diffusiveness}
=\frac{4}{LK^2}\sum_{l=1}^L \sum_{i=1}^{K-1}\sum_{j=i+1}^K (b_{n,i,l} \oplus b_{n,j,l}),
\end{equation}
where $b_{n,l}$ is the $l$-th binary bit of an $L$-bit response from a chip $n$.
This quantity evaluates the degree of the difference among outputs of different RO pairs 
in the same device.
The ideal value of the diffusiveness is 100\% in which all the bit
sequences from different IDs are different.
Figure~\ref{diffusive} shows the diffusiveness as the 
function of the number of RO pairs in the three types of FPGAs.
The results of Fig.~\ref{diffusive} show that 
combination of many RO pairs increases the quality of IDs.

\begin{figure}
\centering
\includegraphics[width=5.7cm]{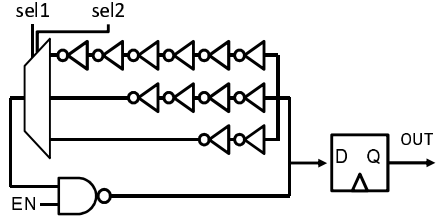}
\caption{
Control of RO frequency by inserting selective 
delay structure. In the present case of FPGAs,
we use primitive LUTs from Altera and Xilinx.
} 
\label{mux}
\end{figure}

The RO frequency can be adjusted by inserting buffers in the RO as shown 
in Fig.~\ref{mux}. Here, the buffers represent the primitive look-up tables (LUTs) prepared by 
Altera and Xilinx. 
It is found that the number of buffers  slightly changes the ID 
in the present case of the automatic routing and placement. 

\begin{table}
\begin{center}
TABLE II\\
{PUF performance measured on the three types of devices}. 
\end{center}
\begin{tabular}{|c|c|c|c|c|}\hline
&  Cyclone V 
&  Spartan-6 
&  Spartan-3E 
&  Ideal Value
\\ \hline
Uniformity 
& 43.54  \%
& 55.66 \%
& 46.74 \%
& 50\%
\\
Reliability
& 98.59 \%
& 96.96 \%
& 97.65 \%
& 100 \%
\\
Uniqueness
& 62.00\%
& 32.52\%
& 42.20\%
& 100 \%
\\ \hline
\end{tabular}
\end{table}

\section{PUF performance evaluation}
We evaluate the PUF performance of our wRO-PUF based 
on the metrics discussed in Ref.\cite{Maiti3,Hori}.
The results of the evaluation are shown in Table II.
Here, uniqueness is calculated from
\begin{equation}
{\rm Uniqueness}=\frac{2}{N(N-1)}\sum_{i=1}^{N-1}\sum_{j=i+1}^N
\frac{HD(R_i,R_j)}{L}\times 100\%,
\end{equation}
where $N$ is the number of the tested devices, $L$ is  
the length of the ID, and $R_i$ is the ID pattern that appears most frequently.
Reliability is calculated from
\begin{equation}
{\rm Reliability}= \left\{1-\frac{1}{T}\sum_{t=1}^{T}\frac{HD(R_i,R_j)}{L} \right\}\times 100\%,
\end{equation}
where $T$ is the number of times sampling is done.
Uniformity is calculated from 
\begin{equation}
{\rm Uniformity}_n=\frac{1}{L}\sum_{l=1}^{L} b_{n,l} \times 100 \%.
\end{equation}
The average results over all samples are listed in Table II.

The uniformity and the reliability are close to the desirable values, 
whereas the uniqueness is smaller than the ideal value.
When we see that the arbiter-PUF shows smaller values of the uniqueness 
from Ref.\cite{Maiti3,Hori}, 
the small value of the uniqueness is considered to
come from the arbiter-PUF performance, 
because our PUF is a combination of the RO-PUF and the arbiter-PUF.

\begin{figure}
\centering
\includegraphics[width=8.5cm]{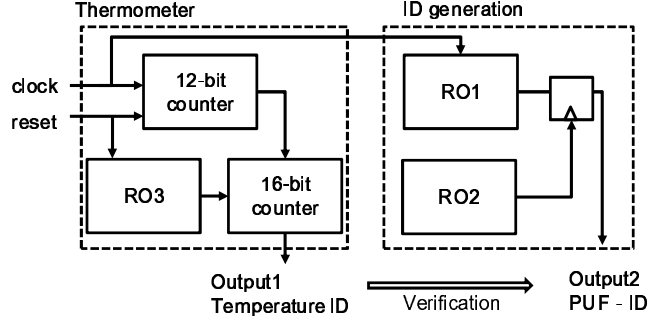}
\caption{
Basic schematics of the wRO-PUF with the thermometer.
The thermometer can detect the change of the environment of wRO-PUF.
Depending on the change of the thermometer (output1), 
we can conjecture the change of PUF-ID (output2).} 
\label{system}
\end{figure}

\begin{figure}
\centering
\includegraphics[width=8.5cm]{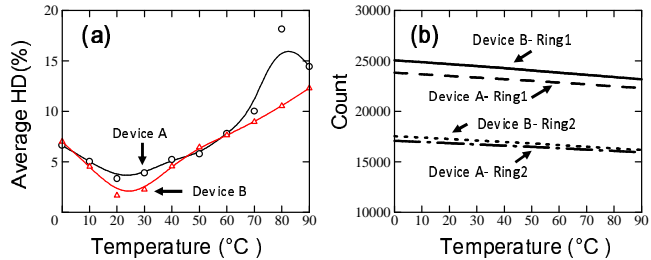}
\caption{
(a) The temperature dependence of 384-bit length 
ID consisting of 12 ROs in Spartan-3E devices.
The average Hamming distance (HD) is calculated 
from the ID at 20~${}^\circ$C.
The ID linearly changes as the temperature difference 
from 20~${}^\circ$C increases. 
(b)
The temperature dependence of the 16-bit counter 
of Fig.~\ref{system} for two ROs of two Spartan-3E devices.
The count linearly decreases as the temperature rises.
The temperature dependence of the count is different 
depending on ROs and devices. This means that the 
temperature measurement can be combined with the 
output of the RO pair.
} 
\label{temperature}
\end{figure}

\section{Temperature and voltage compensations}
In general, as the temperature rises, RO frequency linearly decreases.
Similarly, the change of supplied voltage affects the PUF outputs.
In order to use RO-PUF, we have to take into account 
the effects of these environmental changes. 
The general solution to the changing ID is to add some feedback 
circuit to the PUF system~\cite{Yang}. 
However, for the low-cost PUF and FPGA-PUF,
an analog circuit is not preferable. 

The simple solution to this problem is to detect the environmental change and provide 
IDs that also change according to the environment.
Figure~\ref{system} shows our proposed circuit.
We prepare the 3rd RO, RO3, whose frequency is estimated by 
a counter and shows the status of the environment~\cite{Sayed}.
Figure~\ref{temperature}(a) shows the average HD from the ID 
at 20~${}^\circ$C in the case of sampling done more than thousand times. 
As can be seen, 
the ID tends to change linearly as the temperature difference 
from 20~${}^\circ$C increases.
The small temperature change would be correctable by an error-correcting code. 
Figure~\ref{temperature}(b) shows  the count of the RO3 in Fig.~\ref{system}. 
It can be seen that the count has clear linear dependence on the temperature. 
When the RO3 is closely placed 
near the wRO-PUF, the change of the physical 
environment of wRO-PUF is correctly 
detected by the count of the RO3. Thus, in our scheme, 
by collating the count of RO3 (temperature ID) to the PUF-ID, 
we can check the correct ID of the device.

\section{Conclusion}
We proposed a RO-PUF using 
the initial waveform (wRO-PUF) accompanied by 
a temperature detection system that 
is also constituted by ROs.
Although it is preferable that PUF units are placed compactly in the 
 small area of the FPGA cell to avoid 
extra noises by the global wires in the FPGAs,
the wRO-PUF is implemented just by 
downloading a Verilog circuit, without any manual routing.
We proved the wRO-PUF is applicable to various FPGAs
of Altera and Xilinx.

\section*{Acknowledgment}
TT thanks A. Nishiyama, K. Muraoka, S. Shimizu, 
Y. Komano, T. Marukame, H. Noguchi, T. Kanesige, T. Ootuki,
T. Yamakawa, 
and N. Sakamoto for useful discussion.



\begin{thebibliography}{99}


\bibitem{arbiter} 
D. Lim, J. Lee, B. Gassend, G. Suh, M. van Dijk, and S. Devadas, ``Extracting
secret keys from integrated circuits,” {\it IEEE Trans. Very Large
Scale Integr. (VLSI) Syst}., vol. 13, no. 10, pp. 1200-1205, Oct. 2005.

\bibitem{Lee} 
J.W. Lee, D. Lim, B. Gassend, G.E. Suh, M. van Dijk, S.
Devadas, ``A Technique to Build a Secret Key in Integrated
Circuits for Identification and Authentication Application", in {\it Proc.
Symp. VLSI Circuits},  2004, pp. 176-159.

\bibitem{Guajardo} 
J. Guajardo, S. S. Kumar, G.-J. Schrijen, and P. Tuyls, “FPGA intrinsic
PUFs and their use for IP protection”, in {\it Proc. 9th Int. Workshop on
Cryptographic Hardware and Embedded Systems (CHES’07)}, 2007,
pp. 63-80.

\bibitem{Holcomb}
D. E. Holcomb, W. P. Burleson, and K. Fu, ``Power-up SRAM State as an Identifying Fingerprint and Source of True Random Numbers”, {\it IEEE Trans. Comput.}, vol. 58, no. 9, pp. 1198-120, Sep. 2009. 



\bibitem{Marukame}
T. Marukame, T. Tanamoto, and Y. Mitani,
``Extracting Physically Unclonable Function From Spin Transfer Switching Characteristics in Magnetic Tunnel Junctions", {\it IEEE Trans. Mag}. vol. 50, no. 11,  pp.1-4, Nov. 2014. 

\bibitem{RRAM}
A. Chen,
``Utilizing the Variability of Resistive Random Access Memory to Implement Reconfigurable Physical Unclonable Function", 
{\it IEEE Elec. Dev. Lett.}  vol. 36, no.2 pp. 138-140, Feb. 2015.

\bibitem{Chen}
J. Chen, T. Tanamoto, H. Noguchi, and Y. Mitani,
``Further investigations on traps stabilities in random telegraph signal noise and the application to a novel concept physical unclonable function (PUF) with robust reliabilities", 
{\it 2015 IEEE Symp. VLSI Tech.}  2015, pp.40-41.

\bibitem{Suh} 
G. E. Suh and S. Devadas, “Physical unclonable functions for device
authentication and secret key generation”, in {\it Proc. 44th Ann. Design
Automation Conf. (DAC’07)}, 2007, pp. 9-14.

\bibitem{Maiti0} 
A. Maiti, J. Casarona, L. McHale, and P. Schaumont, “A large scale
characterization of RO-PUF”, in {\it Proc. IEEE Int. Symp. 2010 Hardware-
Oriented Security and Trust (HOST)}, 2010, pp. 94-99.

\bibitem{Maiti1} 
A. Maiti and P. Schaumont, ``Improving the quality of
a physical unclonable function using configurable ring
oscillators", In {\it 19th International Conference on Field
Programmable Logic and Applications 2009
(FPL '09)}, 2009, pp 703-707.


\bibitem{Habib} 
B. Habib , K. Gaj, J.-P. Kaps, “FPGA PUF Based on Programmable
LUT Delays”, in {\it Proc. Euromicro Conf. Digital System Design 2013 (DSD'13)} 
2013, pp 697-704.

\bibitem{Merli} 
D. Merli, F. Stumpf, and C. Eckert, “Improving the quality of ring oscillator
PUFs on FPGAs”, in {\it Proc. 5th Workshop on Embedded
Systems Security(WESS' 10)}, 2010.



\bibitem{butterfly}  
S. Kumar, J. Guajardo, R. Maes, G.-J. Schrijen, and P. Tuyls, “Extended
abstract: The butterfly puf protecting IP on every FPGA”, in
{\it Proc. IEEE Int. Workshop on Hardware-Oriented Security and Trust,
2008 (HOST 2008)}, 2008, pp. 67-70.

\bibitem{FF-PUF}
R. Maes, P. Tuyls, I. Verbauwhede, 
``Intrinsic PUFs from flip-flops on reconfigurable devices", in {\it Proc. Workshop
on Information and System Security (WISSec)}. 2008, pp. 17-26.

\bibitem{Yamamoto} 
D. Yamamoto, K. Sakiyama, M. Iwamoto, K. Ohta, T. Ochiai, M. Takenaka, and
K. Itoh, ``Uniqueness enhancement of puf responses based on the locations of random
outputting rs latches",
in {\it Proc. 13th Int. Workshop on
Cryptographic Hardware and Embedded Systems (CHES’11)}, 2011,
pp. 390-406.

\bibitem{Maiti3}
A Maiti, and P. Schaumont, 
``The Impact of Aging on a Physical Unclonable Function",
{\it IEEE Trans. Very Large
Scale Integr. (VLSI) Syst}, vol 22, no.9, Sept. pp.1854-1864 (2013).

\bibitem{Hori} 
Y. Hori, T. Yoshida, T. Katashita, and A. Satoh, “Quantitative
and statistical performance evaluation of arbiter physical unclonable
functions on FPGAs”, in 
{\it Proc. 2010 IEEE Int. Conf. Reconfigurable Computing and FPGAs (ReConFig2010)}
pp. 298-303.

\bibitem{Yang}
K. Yang, Q. Dong, D. Blaauw, and D. Sylvester,
``A Physically Unclonable Function with BER$<10^{-8}$ for
Robust Chip Authentication Using Oscillator Collapse
in 40nm CMOS",
{\it IEEE Int. Solid-State Circuits Conference  2015}, pp 254-255.


\bibitem{Sayed}
M. A. Sayed, and P. H. Jones, “Characterizing non-Ideal Impacts of Reconfigurable Hardware Workloads on Ring Oscillator-based Thermometers”,in {\it Proc. 2011 IEEE Int. Conf. Reconfigurable Computing and FPGAs (ReConFig2011)}, 2011, pp. 92-98.

\end{thebibliography}
\end{document}